\documentclass[Physsubmission, Phys]{SciPost}

\hypersetup{
    colorlinks,
    linkcolor={red!50!black},
    citecolor={blue!50!black},
    urlcolor={blue!80!black}
}

\usepackage[bitstream-charter]{mathdesign}
\usepackage{lineno}

\usepackage{siunitx}
\DeclareSIUnit\px{px}
\DeclareSIUnit\electron{e}
\DeclareSIUnit\dru{dru}
\DeclareSymbolFont{usualmathcal}{OMS}{cmsy}{m}{n}
\DeclareSymbolFontAlphabet{\mathcal}{usualmathcal}

\begin{document}

\begin{center}{\Large \textbf{
Analysis of radiation damage in silicon charge-coupled devices used for dark matter searches.\\
}}\end{center}

\begin{center}
Steven J. Lee\textsuperscript{1$\star$} on behalf of DAMIC-M collaboration

\end{center}

\begin{center}
{\bf 1} Universit\"{a}t Z\"{u}rich Physik Institut, Z\"{u}rich, Switzerland
\\
*stelee@physik.uzh.ch
\end{center}

\begin{center}
\today
\end{center}


\definecolor{palegray}{gray}{0.95}
\begin{center}
\colorbox{palegray}{
  \begin{tabular}{rr}
  \begin{minipage}{0.1\textwidth}
    \includegraphics[width=30mm]{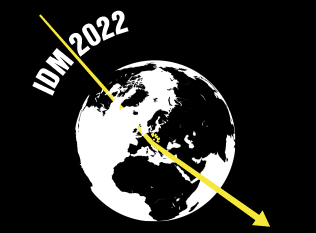}
  \end{minipage}
  &
  \begin{minipage}{0.85\textwidth}
    \begin{center}
    {\it 14th International Conference on Identification of Dark Matter}\\
    {\it Vienna, Austria, 18-22 July 2022} \\
    \doi{10.21468/SciPostPhysProc.?}\\
    \end{center}
  \end{minipage}
\end{tabular}
}
\end{center}

\section*{Abstract}
{\bf
Nuclear recoils in crystal detectors generate radiation damage in the form of crystal defects that can be measured in scientific-grade CCDs as local hot spots of leakage current stimulated by temperature increases in the devices. In this proceeding, we use a neutron source to generate defects in DAMIC-M CCDs, and using increases in leakage current at different temperatures, we demonstrate a procedure for identifying crystal defects in the CCDs of the DAMIC-M experiment.  This is the first time that individual defects generated from nuclear recoils have been studied. This technique could be used to distinguish nuclear recoils from electron recoils in some energy ranges, which would improve the ability of CCD detectors to search for weakly interacting dark matter.
}



\section{Introduction}
\label{sec:intro}
The DAMIC (Dark Matter in CCDs) collaboration has been searching for dark matter using charge-coupled devices (CCDs)\cite{cad}. The CCDs are solid state devices that use the silicon bulk as the ionizing medium. The CCDs are unique compared to the dominant imaging CMOS technology in a way that the CCDs utilize poly-silicon gate structures and field stops to isolate charge packets from adjacent pixels. The advantage of using CCDs as particle detector is that the coupled pixels in the CCDs share a common channel throughout the device and transferring charge packets from one pixel to another has a very high charge transfer efficiency. In addition, even for high substrate voltage the CCD produces very little leakage current making CCDs ideal as high sensitivity particle detectors.

The DAMIC CCDs have a charge transfer inefficiency less than $10^{-4}$\% and leakage current of approximately \SI{2}{\electron}$^-$\SI{}{\per\milli\meter\per\day}\cite{DAMIC:2019dcn}. A modern intervention of DAMIC-CCDs also include skipper amplifiers which utilize floating readout transistor structures allowing non-destructive sampling of charge packets on the readout. By increasing the number of sampling using the skipper amplifier, the DAMIC-M CCDs achieve a readout noise as low as $\sigma_{RMS}=$\SI{0.05}{\electron}$^-$\cite{damicm}. 

In addition to ionizing signals, we can also observe non-ionizing 
events within the CCDs that result in observable lattice damage.  In this proceeding, we report on the observation of radiation damage from the defects caused by single nuclear recoils in DAMIC-M CCDs. The properties of radiation damage allow us to search for signs of particle-lattice interactions which may be used in the search for dark matter.

\section{Lattice defects and dark matter search}
\label{sec:another}
When a foreign particle enters a solid state structure, it loses its energy primarily by Ionizing Energy Loss (IEL) and Non-Ionizing Energy Loss (NIEL) until the particle escapes the solid completely or stops. The total energy loss of the particle is equal to the stopping power of the particle\cite{Kai,moll,personalkai}.
\begin{equation}
\label{stop}
S=S_e+S_n+S_{nr}
\end{equation}
In Equation \ref{stop}, S is the stopping power, $S_e$ is the electronic stopping power equivalent to IEL, $S_n$ is the nuclear stopping power equivalent to NIEL, $S_{nr}$ is the high relativistic term. It is important to note that the NIEL or $S_n$ component damages the lattice structure of the solids\cite{personalkai}.

\subsection{Defect creation}
\begin{figure}[h]
\centering
\begin{tabular}{c c}
\includegraphics[width=0.35\textwidth]{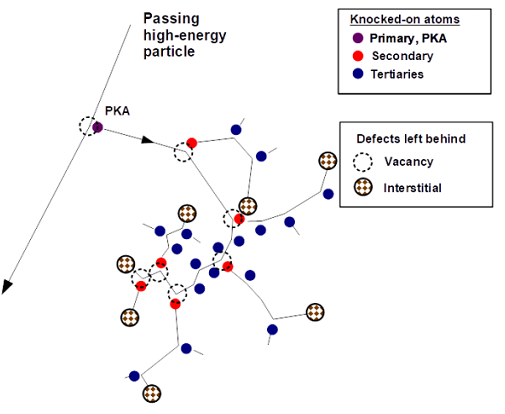} & \includegraphics[width=0.35\textwidth]{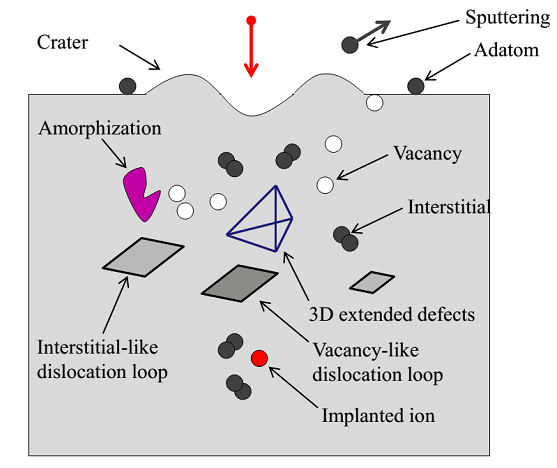}
\end{tabular}
\caption{Types of defects produced in radiation damage\cite{Kai}.}
\label{types}
\end{figure}

In the first few picoseconds after a foreign particle collides with a lattice atom several things happen; the first lattice atom becomes dislodged from its original location (primary knock-on atom, PKA) and knocks out secondary and tertiary atoms forming clusters of defects. Figure \ref{types}(left) demonstrates interaction in a diagram.  An intrinsic silicon crystal has tetrahedral (4 prong) bonds to each other, and as soon as one of the bonds is broken, new bonds are made to the neighbouring atoms. Some of the broken bonds are replaced by Si-Si double or triple bonds, some bonds are made with common impurities such as oxygen, hydrogen, carbon and etc.
Figure \ref{types}(right) outlines the types of defects we can expect from radiation damage. After the first few picoseconds, the lattice is deformed due to these bonds and stabilizes. The stabilized deformations in the lattice structure are called crystal defects. Some semiconductor detectors are intentionally deformed by ion implantation for in-situ or additional doping.

\subsection{Defect identification}
Crystal defects also have a macroscopic electrical property due to the change in the local band-gap resulting from the different allowed energy states from the new lattice bonds. As a result the defective regions can introduce a higher or lower leakage current than the rest of the silicon.

\begin{figure}[h]
\centering
\begin{tabular}{c c}
\includegraphics[width=0.55\textwidth]{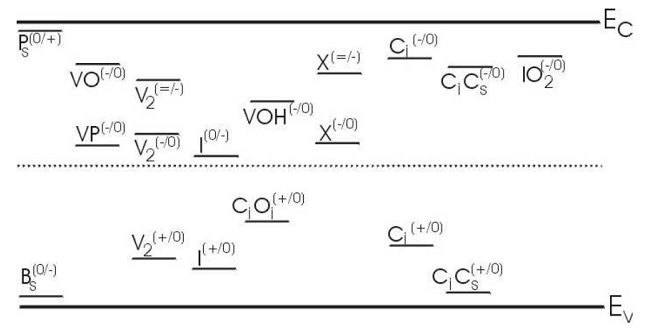} & \includegraphics[width=0.4\textwidth]{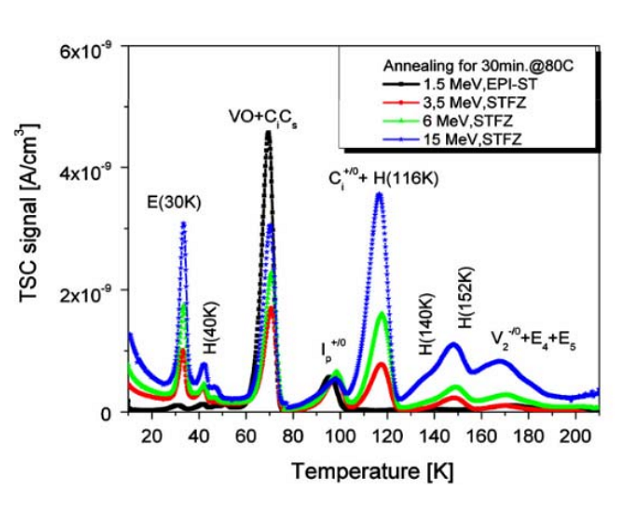}
\end{tabular}
\caption{Defect species in band gap structure\cite{Stahl} and their thermal responses\cite{moll}.}
\label{tscandbandgap}
\end{figure}

The bulk region of the DAMIC-M CCDs are n type and doped with Phosphorous (P). Figure \ref{tscandbandgap} (shows) a sketch of the bandgap structure between the valence band ($E_V$) and conduction band ($E_C$) of silicon. During device production, Boron (B) is typically used for doping silicon positively and Phosphorous is typically used for doping silicon negatively, Vacancy defects (V) can mix with themselves ($V_2$), or with Oxygen (O), Hydrogen (H), Carbon (C), and likewise for Interstitial defects (I). In addition, there are some defects corresponding to a specific bandgap deformation but not yet identified (X). The type of defects produced following a radiation damage is almost entirely stochastic\cite{personalkai}. However all defects have observable macroscopic properties.

Different defect species respond differently to environmental changes such as temperature, bias voltage, and stress. One of the easiest characteristics to test is temperature. Thermally stimulated currents (TSC) is a technique to measure the different temperature response due to different defect species\cite{moll}.

Figure \ref{tscandbandgap} (right) shows an example of the temperature dependence of a TSC signals for four different irradiated samples. The plot shows the abnormal current generation at different temperatures due to the different defect species created.

\section{Neutron Irradiation Campaign}
Typically in high-energy particle physics applications, where the goal is to study performance of irradiated silicon devices, very high energies and fluences are utilized. However the advantages of using the DAMIC-M CCDs are their low energy threshold and the ability to discern individual events. These CCDs can be used to study defect production at the single-event scale.

In the first quarter of 2022 at the University of Washington Center for Experimental Nuclear Physics and Astrophysics(CENPA) CCD test stand, we irradiated a 24 megapixel DAMIC-M CCD (UW6418) using an AmBe source which emits 7400 neutrons per second.

\subsection{Data collection}
All of the data in the experiment were taken in full image mode with 10-minutes exposure followed by 7 minutes of readout time. The CCD temperature was incremented by \SI{5}{\kelvin} ranging from \SI{120}{\kelvin} to \SI{200}{\kelvin}. For consistency, 8 exposures were taken during and after temperature increase. In between each data set, the CCD was warmed to room temperature for reconfiguration inside the test chamber. Table 1 summarizes the 8 data sets of the experiment.

\begin{table}[htbp!]
\begin{tabular}{|c|c|c|c|} \hline
Data set & Type & Source arrangement & Temp. Range(K)\\ \hline
Initial test & Permanent defect search& No source & 120 to 200\\ \hline
Fe-55 Calibration & Pixel value-Fe55 x-ray peaks& Fe-55 & 120 to 200\\ \hline
Proof of Concept & Damage intensity test & Unshielded AmBe & 120 to 200\\ \hline
Before Irradiation 1 & Defect search & No source & 120 to 200\\ \hline
Irradiation 1 & Defect production & Unshielded AmBe & 120 to 160\\ \hline
After Irradiation 1 & Defect search & No source & 120 to 200\\ \hline
Irradiation 2 & Defect production & AmBe & 120 to 200\\ \hline
After Irradiation 2 & Defect search & No source & 120 to 200\\ \hline
\end{tabular}
\caption{Experiments performed with UW6418 DAMIC-M CCDs at CENPA}
\label{table}
\end{table}

The initial test seen in Table \ref{table} was performed without the AmBe source to identify and eliminate defects already present in the CCD prior to neutron irradiation. An $^{55}$Fe Calibration was then performed to calibrate the raw pixel values in Arbitrary Digital Units (ADU) to electron-equivalent energy. A Proof of Concept irradiation was performed with the AmBe source placed less than \SI{2}{\centi\meter} in front of the CCD with minimal lead shielding fir 3 days to confirm the presence of ionization events from the source in the CCD images. Two other irradiations were also carried out: 
\begin{enumerate}
\item Irradiation 1 (5 days): a longer irradiation with the same source configuration, which was halted half way after reaching 160 K due to time constraints.
\item Irradiation 2 (6 days): with the AmBe source placed approximately \SI{11}{\centi\meter} in front of the CCD with \SI{8}{\centi\meter} of lead shielding between the CCD and the source.
\end{enumerate}
Data runs without the source (Before Irradiation 1, After Irradiation 1, After Irradiation 2) were performed after each irradiation to look for newly produced defects.


\subsection{Defect analysis}
The defect analysis was performed by using data collected from initial test, before and after irradiations.

From each data set, the images taken at \SI{200}{\kelvin} were used to locate the repeating patterns. All patterns ranging from single pixels up to vertical saturation lines stretching throughout the CCD were identified as 'tracks' if the edge of the pattern satisfied Equation \ref{grad} in the Appendix.

Only the tracks that appear in more than two images were considered as defects. A typical physical permanent defect unrelated to the irradiation generate significantly larger leakage current and saturates into adjacent pixels. They were easily identified and discarded from the analysis. Most of the defects produced following the neutron irradiation were typically confined to one pixel.

Once the defects have been located, the pixel coordinates were used from all of the data sets. Defects seen from After Irradiation 1 were compared to the defects seen in Before Irradiation 1 to identify newly produced defects. Same analysis was performed After Irradiation 2. As seen in Figure \ref{stsd} the number of newly produced defects increased from irradiation.

In addition, the originating ionization events leading to the defects were identified from the correlation of their spatial coordinates in the images. 
For ionization events with total energy higher than \SI{100}{\kilo\electronvolt}$_{ee}$, nuclear recoils from neutron scattering can be identified by their topology since they generate symmetrical round clusters, unlike the extended tracks or ``worms’’ from electron-recoil backgrounds.
A preliminary analysis shows that at least 80\% of the nuclear recoils identified by topology are spatially correlated with a defect appearing after irradiation.
Lower-energy ionization events are also spatially correlated with defects but, since we cannot tell which events are nuclear recoils, an efficiency estimate is not possible at this time.

We have also been able to categorize the defects by their thermal responses. Depending of the type of defects, similar to the TSC analysis, the defects are stimulated the most at specific temperature. The thermal responses were categorized if at below \SI{190}{\kelvin} the defect satisfies \ref{that}. 
\begin{equation}
\label{that}
z \geq \mu_z^{CCD}+1.5\sigma_z^{CCD}
\end{equation}
Here $z$ is the leakage charge from a single (defective) pixel, and $\mu_z^{CCD}$ is the average leakage charge of the whole CCD and $\sigma_z^{CCD}$ is the standard deviation of the leakage charge of the whole CCD. Using this approach, defects and their response to thermal stimulation was studied.

\section{Conclusion}
By irradiating a silicon CCD with a neutron AmBe source, we have identified for the first time individual defects generated by nuclear recoil events. A preliminary spatial correlation analysis concludes that at least 80\% of recoils with energies >\SI{100}{\kilo\electronvolt}$_{ee}$ produce measurable defects.

\begin{figure}[h]
\centering
\includegraphics[width=0.65\textwidth]{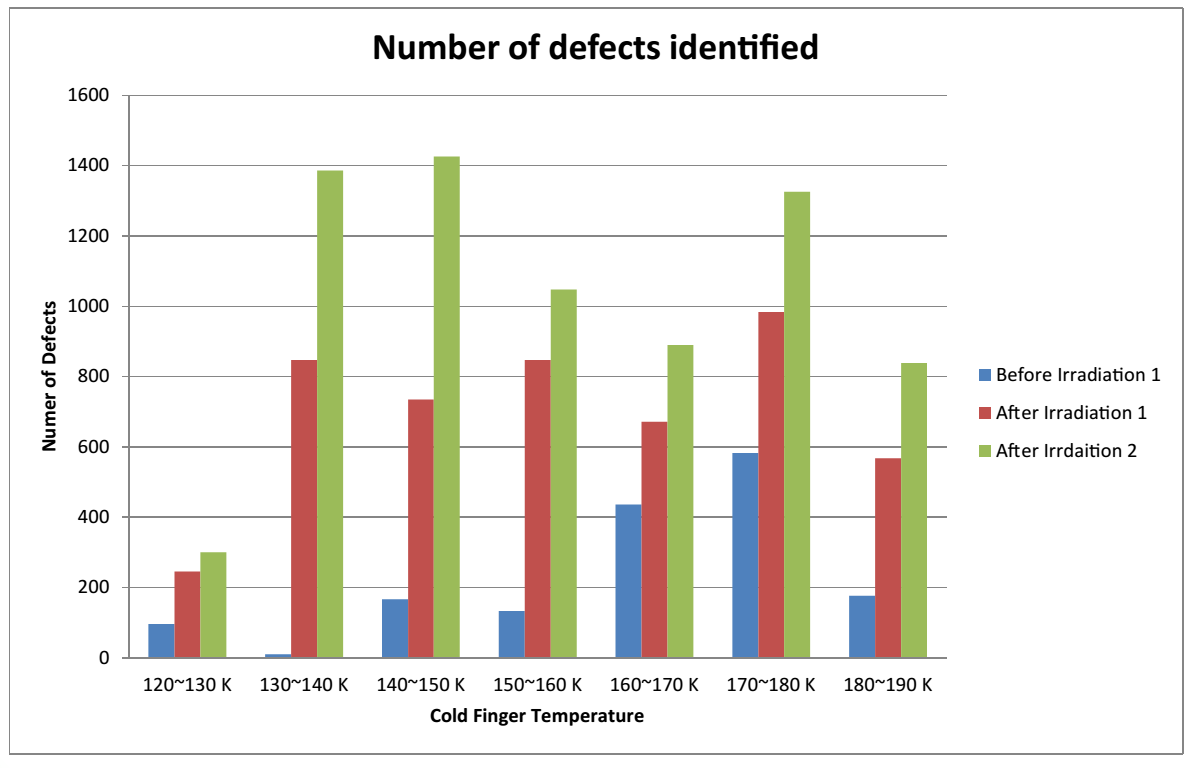}
\caption{Number of defects identified in between neutron irradiations.}
\label{stsd}
\end{figure}

In addition, we have analyzed the thermal stimulation properties of the defects which can be seen in Figure \ref{stsd}.
Figure \ref{stsd} shows the number of defects identified before and after each irradiation. The different colours represent the defects produced prior to and after each irradiation, blue is after the first proof of concept irradiation, red is after the second irradiation and green is after third irradiation. In addition, the defects have the largest gain at outlined cold finger temperatures.

Future work on radiation defects in CCDs is being studied within the RADAC (Radiation Damage in CCDs) group, including the potential application of searching for a daily modulation in the rate of WIMP interactions \cite{fedja}.

The defect analysis is further being developed to search for defects produced from lower recoil energies to look for a low mass DM search using defects.

\section*{Acknowledgements}
\paragraph{RD50}
The author would like to acknowledge M.\,Moll, A.\, Himmerlich, E.C.\,Rivera A.\,Macchiolo, V.\,Subert and Y.\,Gurimskaya for providing insight in searching for radiation damage using TSC and DLTS techniques and providing simulation support for damage spectrum.
\paragraph{Theorists}
The author would like to acknowledge K.\,Nordlund, F.\, Djurabekova, F.\,Kadribasic for providing theoretical insight for damage search and the application of lattice damage in dark matter search.

\paragraph{Funding information}
The author would like to acknowledge the Swiss National Science Foundation (SNSF), the graduate campus of University of Zurich and the University of Zurich Candoc grant (ehemals Forschungskredit) for enabling this research.
\begin{appendix}
\section{Edge gradient search}
\begin{equation}
\left. \frac{dz_{n,m}}{dx_i} \right\rvert_{i=n-2}^{i=n+2} \text{ or } \left. \frac{dz_{n,m}}{dy_j} \right\rvert_{j=m-2}^{j=m+2} \geq \mu_z^{25adj.pix}+1.5\sigma_z^{25adj.pix}
\label{grad}
\end{equation}
In Equation \ref{grad}, $z_{n,m}$ is pixel value (leakage charge collected from an individual pixel) of the pixel index n and m, x is pixel location in horizontal axis of the image, y is pixel number in vertical axis of the image.
\begin{equation}
\mu_z^{25adj.pix} = \frac{\sum_{n-2,m-2}^{n+2,m+2} z_{i,j}}{25}
\end{equation}
 $\mu_z^{25adj.pix}$ is the inclusive average pixel values of 25 surrounding pixels from $i=n-2$ to $i=n+2$ and $j=m-2$ to $j=m+2$. 
\begin{equation}
\sigma_z^{25adj.pix} =  \sqrt{\frac{\sum_{n-2,m-2}^{n+2,m+2} ( z_{i,j} -  \mu_z^{25adj.pix})^2}{25}}
\end{equation}

Similarly $\sigma_z^{25adj.pix}$ is the inclusive standard deviation of the pixel values of 25 surrounding pixels from $i=n-2$ to $i=n+2$ and $j=m-2$ to $j=m+2$.

\end{appendix}
%
%




\begin{thebibliography}{99}

\bibitem{cad}
A.~Aguilar-Arevalo, D.~Amidei, X.~Bertou, D.~Bole, M.~Butner, G.~Cancelo, A.~Casta\~neda V\'asquez, A.~E.~Chavarria, J.~R.~T.~de Mello Neto and S.~Dixon, \textit{et al.}
``Measurement of radioactive contamination in the CCD\textquoteright{}s of the DAMIC experiment,''
J. Phys. Conf. Ser. \textbf{718} (2016) no.4, 042057
doi:10.1088/1742-6596/718/4/042057

\bibitem{DAMIC:2019dcn}
A.~Aguilar-Arevalo \textit{et al.} [DAMIC],
``Constraints on Light Dark Matter Particles Interacting with Electrons from DAMIC at SNOLAB,''
Phys. Rev. Lett. \textbf{123} (2019) no.18, 181802
doi:10.1103/PhysRevLett.123.181802
[arXiv:1907.12628 [astro-ph.CO]].

\bibitem{damicm}
C.~De Dominicis \textit{et al.} [DAMIC-M],
``Simulations and background estimate for the DAMIC-M experiment,''
Nuovo Cim. C \textbf{45} (2021) no.1, 6
doi:10.1393/ncc/i2022-22006-y

\bibitem{Kai}
K. Nordlund and F. Djurabekova, \url{http://www.acclab.helsinki.fi/~knordlun/pub/Nor13apreprint.pdf} 
``Multiscale modelling of irradiation in nanostructures, J. Comput. Electr. 13, 122 (2014),"
invited review paper for special issue on device modelling.

\bibitem{personalkai}
Personal correspondence with Kai Nordland within RADAC.

\bibitem{moll}
Michael Moll, ``Displacement Damage in Silicon Detectors
for High Energy Physics"
IEEE TRANSACTIONS ON NUCLEAR SCIENCE, VOL. 65, NO. 8, AUGUST 2018 1561

\bibitem{Stahl}
J. Stahl, ``Defect Characterisation in High-Purity Silicon after $\gamma$- and Hadron Irraditation." PhD thesis, University of Hamburg, 2004. DESY-THESIS-2004-028, July 2004 (ZEUS).`

\bibitem{fedja}
F.~Kadribasic, N.~Mirabolfathi, K.~Nordlund and F.~Djurabekova,
``Crystal Defects: A Portal To Dark Matter Detection,''
[arXiv:2002.03525 [physics.ins-det]]. 



\end{thebibliography}

\nolinenumbers

\end{document}